\documentclass[twocolumn,floatfix,prb,aps,showpacs]{revtex4}
\usepackage{graphicx}
\usepackage{latexsym}
\usepackage{amsmath}
\begin{document}
\bibliographystyle{apsrev}
\title{String order and adiabatic continuity of
  Haldane chains and band insulators}
\author{F. Anfuso and A. Rosch}
\affiliation{Institute for Theoretical Physics,
University of Cologne, 50937 Cologne, Germany}
\begin{abstract}
  The ground state of spin-1 Haldane chains is characterized by the
  so-called string order. We show that the same hidden order is also
  present in ordinary one-dimensional band insulators. We construct a
  family of Hamiltonians which connects adiabatically band insulators
  to two topologically non-equivalent spin models, the Haldane chain
  and the antiferromagnetic spin-1/2 ladder.  We observe that the
  localized spin-1/2 edge-state characteristic of spin-1 chains is
  smoothly connected to a surface-bound state of band insulators and
  its emergence is not related to any bulk phase transition.
  Furthermore, we show that the string order is absent in any
  dimensions higher than one.
\end{abstract}
\maketitle

The antiferromagnetic spin-1 chain is a prominent example of a
spin-liquid, an exotic state of matter where interactions play a dual
role: on the one hand a strong Hund's rule coupling locally leads to
the formation of magnetic spin-1 degrees of freedom, on the other hand
spin correlations screen magnetism completely. As conjectured by
Haldane in the early eighties \cite{Haldane}, spin-1 chains are
characterized by a finite gap in their spectrum and by exponentially
decaying spin-spin
correlations. Since this seminal work a lot of
progress have been made, theoretically and experimentally, towards a
full description of the so-called ''Haldane phase''.  In
particular, Affleck, Kennedy, Lieb and Tasaki \cite{Affleck} provided a
microscopic understanding of this correlated ground state, devicing an
exactly solvable model (AKLT) adiabatically connected to the spin-1
chain. The AKLT ground state is made up solely of nearest-neighbor
singlets (or valence bonds) and the gap in the system (and therefore
in the spin-1 chain) corresponds to the energy needed to break a bond.

In this paper we want to address the question to what extent the
Haldane chain is different from an ordinary band insulator. The latter has 
also a gap and an even number of electrons per unit
cell and, recently, one of the authors has shown \cite{Achim} that  Luttinger
surfaces cannot be used to distinguish band- from correlated
insulators.  
Two properties, though, are considered peculiar of the Haldane phase:
First, when a spin-1 chain is cut, a localized fractionalized
excitation, namely a spin-1/2, is formed at the boundary of the
chain. These boundary spins have for example been observed in the NMR
profile close to the chain ends of Mg-doped $\rm Y_2BaNiO_5$
\cite{Glarum,Tedoldi}.  Second, the ground state possesses a hidden
long-range order characterizing the entanglement of the spins:  Den
Nijs and Rommelse \cite{Den} noticed that, even though true N\'eel
order is absent in the ground state, any site with $S^{z}=\pm 1$ is 
followed by another with $S^{z}=\mp 1$, separated from the first
by a string of $S^{z}=0$ of arbitrary length. This implies that 
 the so-called string order (SO) parameter 
\begin{equation}
\label{stringorderchain}
SO_{\text{chain}}=\lim_{|i-j|\rightarrow\infty}\Bigl\langle
S^{z}_i\exp\Bigl(i\pi \sum_{l=i+1}^{j-1} S^{z}_l\Bigr) S^{z}_j\Bigr\rangle
\end{equation}
is always finite in the Haldane phase.

\begin{figure}
\includegraphics[width=.50\textwidth,clip]{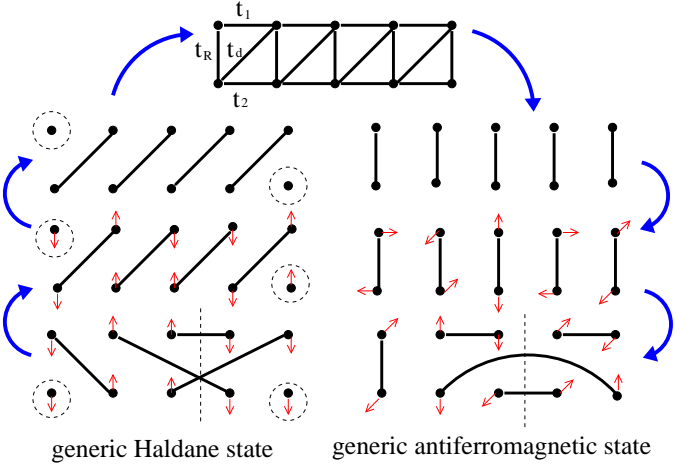}
\caption{Sketch of the adiabatic path connecting a generic Haldane
  state via a band insulator to a generic antiferromagnetic spin
  ladder. 
The two ground states in the lower two rows belong to the distinct topological sectors of
the singlet spin Hilbert space as the number of singlet bonds across
a vertical line (shaded) is either odd or even.} 
\label{string}
\end{figure}

A spin-1 chain can be deviced on a spin-1/2 ladder using a strong
ferromagnetic rung coupling $J_R$ which binds two spin 1/2 into a
single spin 1. In this case the string order parameter takes the form \cite{Watanabe,Nishiyama,White,Kim}
\begin{eqnarray}
SO_{\rm odd}=&-&\lim_{|i-j|\rightarrow\infty}\Bigl\langle
(S^{z}_{1,i}+S^{z}_{2,i})\exp\Bigl(i\pi \sum_{l=i+1}^{j-1}
S^{z}_{1,l}\nonumber \\ &&+S^{z}_{2,l}\Bigr)
(S^{z}_{1,j}+S^{z}_{2,j})\Bigr\rangle \label{soodd}
\end{eqnarray}
and a typical ground state wave function in the Haldane phase is shown
at the bottom of the first column of Fig.~\ref{string}. As emphasized
by Bonesteel \cite{Bonesteel} and Kim {\it et. al.} \cite{Kim}, this
wave function has a topological feature, namely a vertical line
(dashed in the figure) crosses always an {\em odd} number of
singlets. As can be easily seen from the figure, this property is
directly related to the existence of edge states at the boundaries.
On the other hand, an antiferromagnetic rung coupling leads to another
gapful phase, with a ground state in a different topological  sector of
the space of singlet wave functions. Here, always an {\em even} number
of singlets is crossed by a vertical and edge states are absent (see Fig.~\ref{string}). 
Correspondingly \cite{Watanabe,Nishiyama,Kim} the 
''odd'' string order $SO_{\text{odd}}$ vanishes and instead one gets a
finite ''even''  string order $SO_{\text{even}}$
\begin{eqnarray}
SO_{\rm even}=&-&\lim_{|i-j|\rightarrow\infty}\Bigl\langle
(S^{z}_{1,i+1}+S^{z}_{2,i})\exp\Bigl(i\pi \sum_{l=i+1}^{j-1}
S^{z}_{1,l+1}\nonumber\\ & &+S^{z}_{2,l}\Bigr)
(S^{z}_{1,j+1}+S^{z}_{2,j})\Bigr\rangle.
\label{soeven}
\end{eqnarray}
In spin-ladder models these two orders never coexist as they
characterize two different topological sectors.

In the following we will first show that the ground state of an
ordinary band insulator can be adiabatically deformed into the
ground state of either the
Haldane chain or the antiferromagnetic ladder. Then we will
investigate the role both of the string order and of the boundary
states in the band-insulating phase.
Finally, we consider the fate of the string order in dimensions larger
than one.

The starting point of our discussion is the following family of ladder
Hamiltonians (see top of
Fig.~\ref{string}): 
\begin{multline}
H=\sum_{i,\alpha,\sigma} t_\alpha a^{\dagger}_{\alpha,i,\sigma}a_{\alpha,
  i+1,\sigma}+h.c.-\frac{U}{2} n_{\alpha,i,\sigma}\\ + \sum_{i,\sigma} t_R a^{\dagger}_{1,i,\sigma}a_{2,i,\sigma}+
t_D a^{\dagger}_{1,i+1,\sigma}
a_{2,i,\sigma}+h.c.\\+ U \sum_{i,\alpha}
n_{\alpha,i,\uparrow}n_{\alpha,i,\downarrow}
+J_R \sum_i{\bf S}_{1,i}{\bf S}_{2,i}
 \label{hamiltonian}
\end{multline}
where  $\alpha$ indicates the two legs of the ladder 
(or  two orbitals of an atom). 
Varying the set of parameters $[t_1,t_2,t_R,t_D,U,J_R]$, it is possible to
span a rich phase diagram including a Mott-insulating Haldane phase
($U\gg t_i$, $J_R+t_R^2/U<0$), an antiferromagnetic ladder ($U\gg
t_i$, $J_R= 0$) and band-insulators ($U=0, J_R=0$).
We now describe how these states can be connected adiabatically
following the route depicted in Fig.~\ref{string}. First, it has been
shown both numerically \cite{White,Watanabe,Nishiyama} and using analytic arguments 
\cite{Shelton,Kim} that the
Haldane chain can be
adiabatically deformed into a spin-1/2 model with only diagonal antiferromagnetic
couplings, and, similarly, the antiferromagnetic ladder is connected
to a pure rung singlet model (second last row of Fig.~\ref{string}).
These local antiferromagnets can be realized as the large $U$ limit
of the two-sites Hubbard Hamiltonian
\begin{multline}
H_{\rm
  loc}=\sum_{\sigma}t
(c_{1,\sigma}^{\dagger}c_{2,\sigma}+h.c.)-\frac{U}{2}(n_{1,\sigma} +n_{2,\sigma})\\ + U(n_{1,\uparrow}n_{1,\downarrow}
+n_{2,\uparrow}n_{2,\downarrow})
\end{multline}
for the diagonal or the rung, respectively.
For arbitrary $U/t$, the three lowest eigenvalues of $H_{\rm loc}$ are 
a singlet with energy $E_s= -\frac{1}{2}(U+\sqrt{U^2+16 t^2})$, 
 a triplet
 with  $E_t = -U$ and a charge excitation with
$E_c=-\frac{U}{2}-|t|$. The gap
\begin{eqnarray}
\Delta &=& \frac{1}{2} \min\!\left[\sqrt{U^2+16 t^2}-U,
\sqrt{U^2+16 t^2}-2 |t| \right]
\end{eqnarray}
is always finite and, therefore, one can switch off the
interaction $U$ completely, connecting adiabatically
the Mott- and the band-insulating ground states. Note that the nature of the lowest excitation
changes completely when $E_c=E_1$ while the ground state rotates
smoothly \cite{Achim} from a state describing two localized spins to one where two
electrons occupy the bonding band.

\begin{figure}
\includegraphics[width=.45\textwidth,clip]{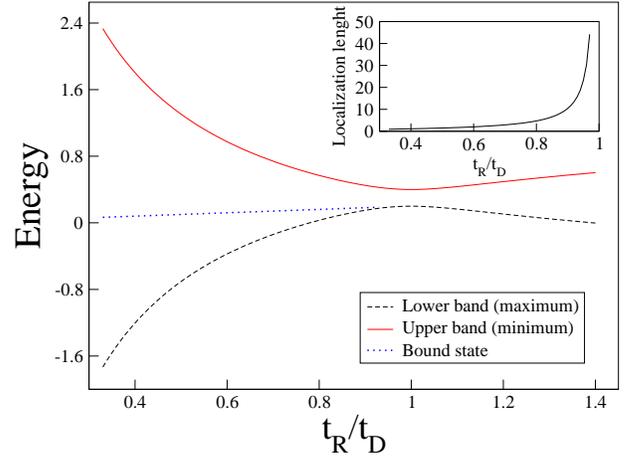}
\caption{Energy of the bound state compared to the maximum
  (minimum) of the upper (lower) band as a function of $t_R/t_D$ 
  ($t_1=-0.1$, $t_2=-0.2$ and $t_R=-1$). In the inset the localization
  length of the bound state is shown.}
\label{bondstate}
\end{figure}
As a final step, we have to show that these two local insulators can
be adiabatically connected in the non-interacting limit, $U,J_R=0$
where the band-structure is given by 
\begin{eqnarray}
E^{1,2}&=&(t_1+t_2)\cos(ka)\pm[(t_1-t_2)^2\cos^2(ka)+\nonumber\\ &&(t_R-t_D)^2+2t_Rt_D(1+\cos(ka))]^{1/2}.
\end{eqnarray}
The states in the second row of Fig.~\ref{string} are described by
$t_1, t_2, t_R=0$ and $t_1, t_2, t_D=0$, respectively.
For $2 |t_1+t_2|<|t_D+t_R|+|t_D-t_R|$ and $t_1 \neq t_2$, the state is
always a band insulator with a finite gap. Therefore by reducing $t_D$
and increasing $t_R$ for finite $t_1-t_2$ we can complete the
adiabatic path of Fig.~\ref{string}.
 For $t_1=t_2$ the Hamiltonian of
Eq.~(\ref{hamiltonian}) has a 
quantum critical point at $t_R=t_D$ (related to an extra particle-hole
symmetry of this model). In Fig.~\ref{bondstate} we show
that for small $t_1-t_2$ the gap remains finite upon tuning from the
diagonal to the rung insulator.

\begin{figure}
\includegraphics[width=.40\textwidth,clip]{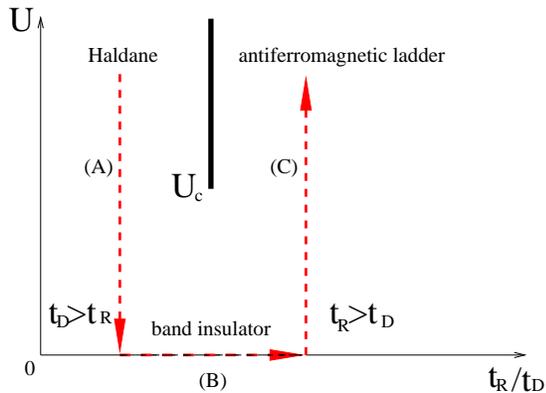}
\caption{Schematic phase diagram of the generalized Hubbard ladder
  (\ref{hamiltonian}). The dashed lines denote the adiabatic path of
  Fig.~\ref{string}. For sufficiently large interactions, either a
  second or first  order transition or new intervening phases are expected.}
\label{adipath}
\end{figure}
After having shown the adiabatic continuity, we consider now the
evolution of the string order along the adiabatic path. 
 In Eqns.~(\ref{soodd}) and (\ref{soeven})  the $S^z_{1-2,i}$ were
 operators living in the spin Hilbert space while for the Fermionic model
 we have to replace those by 
$S^z_{1-2,i}=\frac{1}{2}(n_{1-2,i,\uparrow}-n_{1-2,i,\downarrow})$. 
For the two local singlet states of the third row of Fig.~\ref{string} the
string order is maximal \cite{Kim}, with
$SO_{\rm odd}=\frac{1}{4}$, $SO_{\rm even}=0$ and $SO_{\rm
  even}=\frac{1}{4}$, $SO_{\rm odd}=0$, respectively.
Remarkably, the string order is also finite for the two local band
insulators of the second row: $SO_{\rm odd}=\frac{1}{16}$, $SO_{\rm even}=0$ and $SO_{\rm
  even}=\frac{1}{16}$, $SO_{\rm odd}=0$, respectively. The factor $4$
reduction arises because empty and doubly occupied states appear with
a probability of $1/2$, affecting the spin operators at the beginning
and at the end of the string. The string, however, still contributes with a
factor $1$. How can these different ''order parameters'' be connected?
To answer this question we calculate the two string orders along the
adiabatic path connecting the two local insulators (i.e. on segment B 
of Fig.~\ref{adipath}). 
With the help of the identities $e^{\alpha
  c^{\dagger}c}=:e^{(e^{\alpha}-1)c^{\dagger}c}:$ and 
$S^z_i=-\frac{1}{4}\bigl(:e^{-2c^{\dagger}_{i,\uparrow}c_{i,\uparrow}}:-:e^{-2c^{\dagger}_{i,\downarrow}c_{i,\downarrow}}:\bigr)$
where $:...:$ denotes normal ordering,
we can reformulate Eq.~(\ref{soodd}) as 
\begin{eqnarray}
SO_{\rm
  odd}(i,j)&=&\langle\frac{1}{16}:\sum_{r,l=1,2}(e^{-2n_{r,i,\uparrow}}-e^{-2n_{r,i,\downarrow}})
\nonumber\\ & \times & \exp\Bigl (\sum_{\begin{subarray}{l}k=i+1\\s=1,2\end{subarray}}^{k=j-1}
(-1+i)n_{s,k,\uparrow}+(-1-i)n_{s,k,\downarrow}\Bigr )\nonumber\\
 & \times &(e^{-2n_{l,j,\uparrow}}-e^{-2n_{l,j,\downarrow}}):\rangle
\label{norder}
\end{eqnarray}
and a similar expression is valid for  Eq.~(\ref{soeven}).
In analogy to the string order in the spin sector one can also
introduce a similar quantity in the charge sector, see Appendix~\ref{chargeorder}.

 For a
non-interacting system, one can easily evaluate expectation values of
the form $\langle :e^{c^\dagger_i A_{ij} c_j}: \rangle = 
{\text{Det}}\!\left[1+G\cdot A\right]$ using functional integration,
where $G$ is the equal-time Greens function matrix. We therefore obtain
\begin{equation}
SO^z_{\rm
  e/o}(i,j)=\frac{1}{16}\sum_{\begin{subarray}{l}\alpha,\beta=\uparrow,\downarrow\\r,k=1,2\end{subarray}}(-1)^{2+\delta^{\alpha,\beta}} \rm Det \it
\bigl[I+\emph{G}\cdot A_{e/o}
^{r,i,\alpha,k,j,\beta}\bigr],
\label{determinant}
\end{equation} 
where the matrices $A_{e/o}$ can be read off from Eq.~(\ref{norder}) 
and the $G$ matrix can easily be obtained numerically for arbitrary
$t_1, t_2, t_R, t_D$. 
Also the determinant in Eq.~(\ref{determinant}) can be easily computed 
 for  large systems and the SO parameters are obtained from the
limit $|i-j|\gg \xi$, where $\xi\propto 1/\Delta$ is the correlation length. 
\begin{figure}
\includegraphics[width=.45\textwidth,clip]{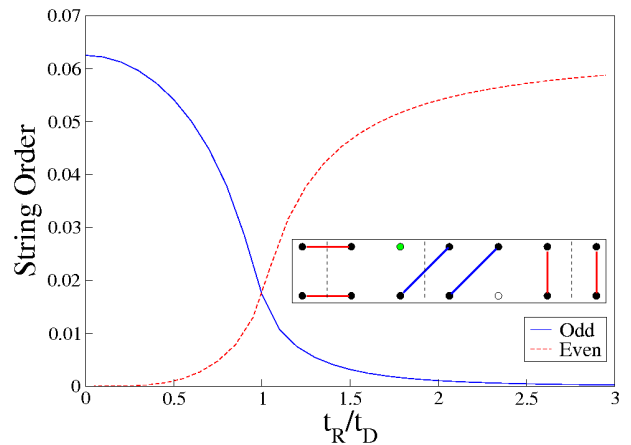}
\caption{The even and odd string orders, $SO_{\rm even}$ and $SO_{\rm
    odd}$, for the non-interacting ladder as a function of $t_R/t_D$
  ($t_1=-0.1$, $t_2=-0.2$, $t_R=-1$). Inset: in the presence of charge
fluctuations a vertical line can cross both even and odd number of
singlets. Accordingly the two string orders can coexist.}
\label{stringorder}
\end{figure}

We find that in one-dimensional band insulators string order is always
present -- even in the absence of interactions!  In contrast to the
pure spin models generically both string order parameters are finite
(see Fig.~\ref{stringorder}) and they are smooth functions of
$t_R/t_D$ for $t_1\ne t_2$. In our model, for $t_R=t_D$ the two string
orders are equal by symmetry, $SO_{\rm even}=SO_{\rm odd}$ (this 
follows from the more general relation
$SO_{\rm even}(t_R,t_D,t_1,t_2)=SO_{\rm odd}(t_D,t_R,t_2,t_1)$
that can be understood in terms of the following transformation on
the ladder: shift of the lower leg by one lattice site and
mirror-reflection of 
the ladder respect the x-axis). Only for
$t_1=t_2$ and $t_R=t_D$, when the gap vanishes, both $SO_{\rm even}$
and $SO_{\rm odd}$ deplete to 0 because at the critical point singlets
of arbitrary length are generated.  The two orders, $SO_{\rm odd}$ and
$SO_{\rm even}$, coexist in gapped insulators also in the presence of
strong interactions as long as (virtual) charge fluctuations are
allowed. This can be seen from the inset of Fig.~\ref{stringorder}: a
doubly occupied or empty site allows to switch from diagonal to rung
singlets.


We now outline the consequences of our discussion for the phase
diagram of generalized Hubbard ladders, sketched in
Fig.~\ref{adipath}. First, gapped systems are stable against small
perturbations and therefore there is a finite range of parameters
which allows to connect adiabatically Haldane states and
antiferromagnetic ladders. Second, in a pure spin model it is {\em
  not} possible to connect these two phases as their wave functions
live in different topological sectors of the spin Hilbert space.
Therefore, we expect that beyond some critical value of $U$ a phase
transition line (first or second order) or an intervening phase should
separate the even and odd sectors\cite{Kim,Balents,Fabrizio,Kolezhuk}.

For pure spin models, Starykh and Balents\cite{Balents} analyzed the
possible intervening phases and concluded that without extra
fine-tuning there is never a direct second order transition from the
Haldane to the antiferromagnetic ladder.  Either the transition is
first order or intervening phases with broken translational invariance
appear. The absence of a direct second-order transition is consistent with
our finding that the two high-symmetry phases are adiabatically
connected. 

\begin{figure}
\includegraphics[width=.45\textwidth,clip]{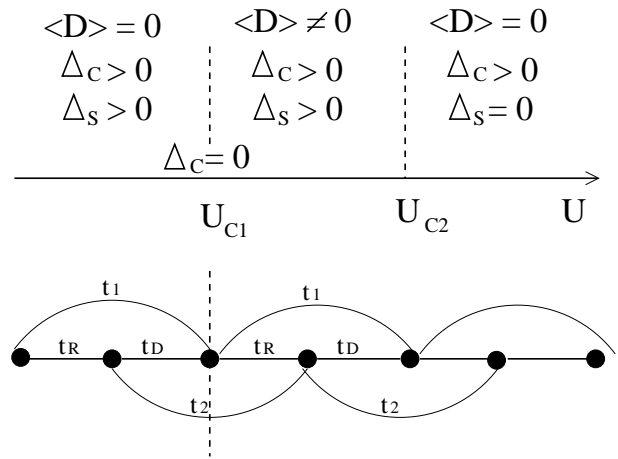}
\caption{Top: the three different phases of the ionic Hubbard model as
a function of $U$. $\Delta_c$ and $\Delta_s$ are the charge and spin
gaps and $D=\frac{1}{L}\sum_{j,\sigma}(-1)^j\langle c^{\dagger}_{j+1,\sigma}c_{j,\sigma}+
c^{\dagger}_{j,\sigma}c_{j+1,\sigma}\rangle$ is the dimerization order
parameter \cite{Fabrizio,Noack}.
Bottom: the ladder of Eq.~(\ref{hamiltonian}) unfolded on a chain. 
An extra mirror symmetry is present for $t_R=t_D$ and $J_R=0$.}
\label{Gaps}
\end{figure}

To gain further insight in nature of the phase diagram sketched in
Fig.~\ref{adipath} at finite $U$, we observe that for $t_R=t_D$ and
$J_R=0$, our ladder gains an extra mirror symmetry (this is evident
unfolding the ladder on a chain, as shown in the lower part of
Fig.~\ref{Gaps}) and we expect that its properties are similar to the
ones of the well-studied\cite{Nagaosa,Fabrizio,Noack,Japaridze}
one-dimensional {\em ionic} Hubbard model (i.e. a Hubbard model with a
staggered potential) which has the same symmetries. For this system
upon increasing $U$ (see upper part of Fig.~\ref{Gaps}) first the
charge gap of the band insulator closes and one obtains a dimerized
phase for $U_{c1}<U<U_{c2}$.  For $U>U_{c2}$ the system becomes a
uniform, gappless Mott insulator (effectively a uniform Heisenberg
{\em chain}\cite{Nagaosa}). We therefore expect that $U_c$ in the
schematic phase diagram of Fig.~\ref{adipath} has a similar role as
$U_{c1}$, i.e. the charge gap vanishes and a dimerized phase (not
shown in Fig.~\ref{adipath}) appears for $U>U_{c1}$, $t_R \sim t_D$,
where $SO_{\rm even}$ is bigger or smaller then $SO_{\rm odd}$,
dependingly on which of the two degenerate ground states one
considers.  According to the results of Starykh and
Balents\cite{Balents} discussed above, we do, however, not expect a
gapless phase for generic parameters (e.g. $J_R >0$). The role of
string order in the charge sector of the ionic Hubbard model is
briefly discussed in Appendix~\ref{chargeorder}.

At this point we would like to investigate the emergence of the
edge-states following again the path depicted in Fig.~\ref{string}.
The localized spin 1/2 at the boundary of the Haldane phase evolves
smoothly in a surface bound state of the ''diagonal'' band insulator.
For finite interactions and a proper choice of the chemical potential,
this surface bound state will be singly occupied giving rise to a
localized spin.  In Fig.~\ref{bondstate} we show how this edge state
merges with one of the bands when $t_R$ is increased for $t_1\neq
t_2$. At this point, the localization length of this state diverges
with $(t_R/t_D-const.)^{-1}$ (inset of Fig.~\ref{bondstate}). Nevertheless,
this is not related to any bulk phase transition.

\begin{figure}
\includegraphics[width=.45\textwidth,clip]{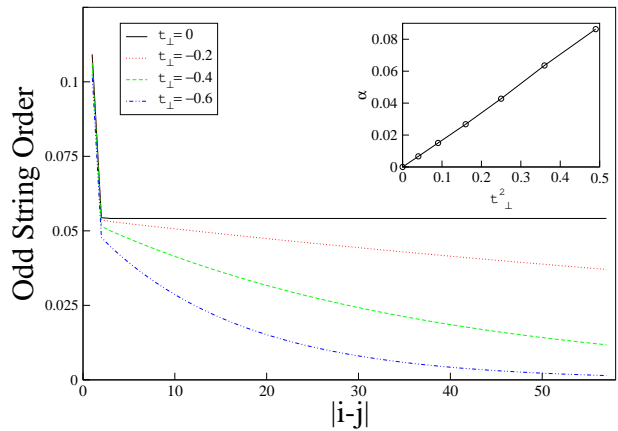}
\caption{The odd string order of Eq.~(\ref{determinant}) for a two-dimensional
  lattice decays as $SO_{\rm odd}\approx e^{-\alpha |i-j|}$
 (with $t_1=-0.1$,
  $t_2=-0.2$, $t_R=-1$, $t_D=-2$ and $t_\perp=0,-0.2,-0.4,-0.6$ from
  up to down). Inset: $\alpha$ as a function of $t_{\perp}^2$.}
\label{2dstring}
\end{figure}

Up to now we have only discussed one-dimensional ladders and it is
interesting to study the role of string order in higher dimensions.
Especially in gapped systems one would naively expect that any type of
long-range order is stable against small perturbations like a coupling
to neighboring ladders. Using the methods described above, we can
easily calculate the string order for two- and three-dimensional band
insulators numerically. Surprisingly we find that for arbitrarily weak
inter-ladders coupling, $t_\perp$, the string order decays
exponentially, $SO(i,j) \sim e^{-\alpha |i-j|}$ with $\alpha \propto
t_\perp^2$ (see Fig.~\ref{2dstring}). A similar observation has been made previously by Todo
{\it et al.} \cite{Todo}, who studied spin-1 ladders numerically.  The
decay of the string order is a consequence of (rare) inter-ladders
singlets which introduce a random phase $e^{\pm i \pi/2}$ in the string.  We note
that obvious generalizations \cite{Todo} of the string order parameter
to higher dimensions (replacing strings by squares or cubes) are
ineffective again due to ''dangling singlets'' at the infinite surface
of such structures.

In conclusions, we have shown that three phases with both spin- and
charge gaps and two electrons per unit cell, namely Haldane chains,
band-insulators and antiferromagnetic ladders, are actually all the
same in the sense that their unique ground states can be adiabatically
connected (the excitation spectra, however, may change). This
cousinship is reflected in the prevalence of string order in all these
one-dimensional phases, even in non-interacting band insulators.
However, string order is not robust in higher dimensions, where it is
destroyed by arbitrarily weak coupling\cite{Anfuso}.  An interesting
open question is the precise role of the Schrieffer-Wolff
transformation, a central tool to derive effective models. Such a
unitary transformation eliminates charge degrees of freedom completely
and therefore maps the non-magnetic ground state of Hamiltonians with
charge gap into one of the topological sectors of the spin models.  As
we have connected two {\em different} sectors adiabatically along our
path, this seems to imply that the Schrieffer-Wolff transformation has
to break down at some point. This is consistent with a scenario where
at $U_c$ in Fig.~\ref{adipath} the charge gap closes. However, one
also has to take into account that a Schrieffer-Wolff transformation
is not uniquely defined as it depends e.g. on the choice of a
single-particle basis and therefore the topological sector may not be
uniquely defined in the presence of charge fluctuations.

The rapid progress in the control of fermionic atoms
in optical lattices will make  possible to study the adiabatic
evolution of correlated ground states also experimentally in the near future
\cite{Kohl,Diener}.

We acknowledge useful discussions with 
R.~Moessner, E.~M\"uller-Hartmann, A.A.~Nersesyan, A.~Schadschneider, A.M.~Tsvelik and
 J.~Zaanen, J.~Zittartz and, especially, G.I.~Japaridze. We thank for
 financial support of the DFG under SFB 608.

\begin{appendix} 
\section{Charge string order in band- and Mott-insulators}\label{chargeorder} 
In close analogy with the case of 
the spins (Eqns.~(\ref{soodd}) and
(\ref{soeven}) in the text), we can construct string order parameters
also in
the charge sector.
We define $SO^c_{\rm even}$ and $SO^c_{\rm odd}$ as
\begin{eqnarray}
SO^c_{\rm odd}=& &\lim_{|i-j|\rightarrow\infty}\Bigl\langle
(\delta n_{1,i}+\delta n_{2,i})\exp\Bigl(i\frac{\pi}{2} \sum_{l=i+1}^{j-1}
\delta n_{1,l}\nonumber \\ &&+\delta n_{2,l}\Bigr)
(\delta n_{1,j}+\delta n_{2,j})\Bigr\rangle \label{sooddcharge}\\
SO^c_{\rm even}=& &\lim_{|i-j|\rightarrow\infty}\Bigl\langle
(\delta n_{1,i+1}+\delta n_{2,i})\exp\Bigl(i\frac{\pi}{2} \sum_{l=i+1}^{j-1}
\delta n_{1,l+1}\nonumber\\ & &+\delta n_{2,l}\Bigr)
(\delta n_{1,j+1}+\delta n_{2,j})\Bigr\rangle
\label{soevencharge}
\end{eqnarray}
where $\delta n_i=n_{i,\uparrow}+n_{i,\downarrow}-\langle
n_{i,\uparrow}+n_{i,\downarrow}\rangle$.

\begin{figure}
\includegraphics[width=.45\textwidth,clip]{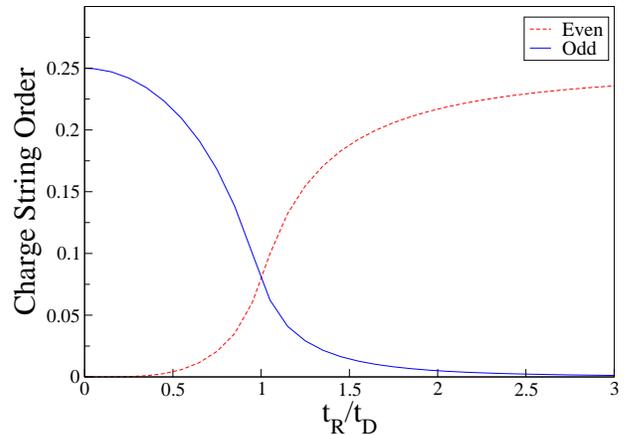}
\caption{$SO^c_{\rm odd}$ and
$SO^c_{\rm even}$ as a function of $t_R/t_D$ for the set of parameters
$t_1=-0.1$,
$t_2=-0.2$ and $t_R=-1$. }
\label{chargestring}
\end{figure}

As shown in Fig.~\ref{chargestring}, both $SO^c_{\rm odd}$ and
$SO^c_{\rm even}$ are finite for a generic band insulator and - up to
a trivial factor $1/4$ - very similar (but not identical) to the spin
string order parameters of Fig.~\ref{stringorder}.
In the
purely diagonal ($t_R=0$) and vertical ($t_D=0$) limits 
only one of the two string orders is different from 0.   
\vspace{10pt}

The presence of charge string order comes as a natural consequence of the
symmetry existing in the one-dimensional band insulator between
the spin and the charge
sectors. In bosonization language, both the spin and charge fields
are pinned due to the presence of a relevant
cosine term in the low energy Bosonic Hamiltonian \cite{Gogolin},
leading therefore to finite string order parameters \cite{Kim,Anfuso}.
As already discussed in the main text, this scenario is stable also in presence of
interactions (at least for truly one-dimensional systems), as long as both the Bosonic fields are massive. 
In the limit of large $U$, the charge string orders will be 
be strongly suppressed by powers of  $1/U$ due to the smallness of the charge
fluctuations (becoming strictly zero for a pure spin model,
i.e. $U=\infty$) but will remain finite for any finite $U$.
 
In the case of the ionic Hubbard model (corresponding to the higher
symmetry manifold $t_R=t_D$ and $J_R=0$ in the parameter space of
Eq.~(\ref{hamiltonian})) the charge field becomes massless at
$U=U_{c1}$ due to the competition between two relevant cosine
perturbations \cite{Fabrizio} (the on-site energy and the Coulomb
interaction). As both for $U<U_{c1}$ and  $U>U_{c1}$ the charge field
is locked, the charge string order will be finite but it will vanish at the
gapless point $U=U_{c1}$ (the spin string order stays finite at $U_{c1}$). 

\end{appendix}


\begin{thebibliography}{999}

\bibitem{Haldane} F.D.M.~Haldane, { Phys.~Rev.~Lett.}~{\bf 50}, 1152
  (1983); { Phys.~Lett.}~{\bf 93A}, 464 (1983).

\bibitem{Affleck} I.~Affleck, {\it et al.}, { Phys.~Rev.~Lett.}~{\bf
    59}, 799 (1987); { Comm.~Math.~Phys.}~{\bf
    115}, 477 (1988).

\bibitem{Achim} A.~Rosch, preprint cond-mat/0602656.

\bibitem{Glarum} S.H.~Glarum, {\it et al.},{ Phys.~Rev.~Lett.}~{\bf 67}, 1614 (1991).

\bibitem{Tedoldi} F.~Tedoldi, R.~Santachiara and M.~Horvati\'c, { Phys.~Rev.~Lett.}~{\bf 83}, 412 (1999).

\bibitem{Den} M.P.M.~Den Nijs and K.~Rommelse, { Phys.~Rev.~B} {\bf 40}, 4709 (1989).

\bibitem{Kim} E.H.~Kim,  {\it et al.}, { Phys.~Rev.~B} {\bf 62}, 14965 (2000).

\bibitem{White} S.R.~White, Phys. Rev. B {\bf 53}, 52 (1996).

\bibitem{Watanabe} H.~Watanabe, Phys. Rev. B {\bf 52}, 12508 (1995).

\bibitem{Nishiyama} Y.~Nishiyama, N.~Hatano and M.~Suzuki,
  J. Phys. Soc. Jpn. {\bf 64}, 1967 (1995).

\bibitem{Bonesteel} N.E.~Bonesteel, { Phys.~Rev.~B} {\bf 40}, 8954 (1989).

\bibitem{Shelton} D.G.~Shelton, A.A.~Nersesyan and A.M.~Tsvelik, { Phys.~Rev.~B} {\bf 53}, 8521 (1996).

\bibitem{Balents} O.A.~Starykh, L.~Balents, {Phys.~Rev.~Lett.} {\bf 93}, 127202 (1997).
\bibitem{Kolezhuk} A.K.~Kolezhuk, H.-J~Mikeska, { Phys.~Rev.~B} {\bf 56}, 11380 (1997).
\bibitem{Fabrizio} M.~Fabrizio, A.O.~Gogolin and A.A.~Nersesyan, {Phys.~Rev.~Lett.} {\bf 83}, 2014 (1999).

\bibitem{Noack} S.R.~Manmana, V.~Meden, R.M.~Noack, K.~Sch\"onhammer,{ Phys.~Rev.~B} {\bf 70}, 155115 (2004).

\bibitem{Japaridze} A.P.~Kampf, M.~Sekania, G.I.~Japaridze, Ph.~Brune,
  {J.~Phys.~Cond.~Mat.} {15}, 5895 (2003).

\bibitem{Nagaosa} N.~Nagaosa and J.~Takimoto,  J. Phys. Soc. Jpn. {\bf 55}, 2735 (1986).

\bibitem{Todo} S.~Todo {\it et al.}, { Phys.~Rev.~B} {\bf 64}, 224412 (2001).

\bibitem{Anfuso} F.~Anfuso, A.~Rosch, pre-print cond-mat/0702064. 

\bibitem{Kohl} M.~K\"ohl {\it et al.}, { Phys.~Rev.~Lett.}~{\bf 94}, 080403 (2005).
\bibitem{Diener} R.B.~Diener and T.L.~Ho, { Phys.~Rev.~Lett.}~{\bf
    96}, 010402 (2006).
\bibitem{Gogolin} A.O.~Gogolin, A.A.~Nersesyan, A.M.~Tsvelik,
  \emph{Bosonization of Strongly Correlated Systems }, Cambridge (1998).



\end{thebibliography}
\end{document}